\documentclass[
reprint,
superscriptaddress,
nofootinbib,
amsmath,
amssymb,
aps,
prd,
]{revtex4-2}

\usepackage{graphicx}
\usepackage{dcolumn}
\usepackage{xcolor}
\usepackage{subfigure}
\usepackage{bm}
\usepackage{amsmath}
\usepackage{makecell}
\usepackage{hyperref}

\begin{document}

\preprint{APS/123-QED}

\title{Nonlinear Evolution of unstable Charged de Sitter Black Holes with Hyperboloidal Formalism}

\author{Zhen-Tao He}
\email{hezhentao22@mails.ucas.ac.cn}
\affiliation{\textit{School of Physical Sciences, University of Chinese Academy of Sciences, Beijing 100049, China}}

\author{Qian Chen}
\email{chenqian192@mails.ucas.ac.cn}
\thanks{Corresponding author}
\affiliation{\textit{School of Fundamental Physics and Mathematical Sciences, Hangzhou Institute for Advanced Study, University of Chinese Academy of Sciences, Hangzhou, Zhejiang 310024, China}}

\author{Yu Tian}
\email{ytian@ucas.ac.cn}
\thanks{Corresponding author}
\affiliation{\textit{School of Physical Sciences, University of Chinese Academy of Sciences, Beijing 100049, China}}
\affiliation{\textit{Institute of Theoretical Physics, Chinese Academy of Sciences, Beijing 100190, China}}

\author{Cheng-Yong Zhang}
\email{zhangcy@email.jnu.edu.cn}
\thanks{Corresponding author}
\affiliation{\textit{Department of Physics and Siyuan Laboratory, Jinan University, Guangzhou 510632, China}}

\author{Hongbao Zhang}
\email{hongbaozhang@bnu.edu.cn}
\thanks{Corresponding author}
\affiliation{School of Physics and Astronomy, Beijing Normal University, Beijing 100875, China}


\date{\today}

\begin{abstract}
Based on the hyperboloidal framework, we research the dynamical process of charged de Sitter black holes scattered by a charged scalar field.
From the linear perturbation analysis, with the coupling strength within a critical interval, the charged scalar field with a superradiance frequency can induce the instability of the system.
To reveal the real-time dynamics of such an instability, the nonlinear numerical simulation is implemented.
The results show that the scalar field grows exponentially in the early stages and drastically extracts the charge from the black hole due to the superradiance, analogous to the charged black hole in a closed system.
Differently, after saturation, the scalar field can not coexist with the central black hole stably and dissipates beyond the cosmological horizon slowly, leaving behind a bald black hole.
\end{abstract}

\maketitle


\section{Introduction}
Superradiance, a radiation amplification process, is of great significance in numerous fields of physics, particularly in relativity and astrophysics \cite{Brito2020}. 
In General Relativity, black hole (BH) superradiance results from the interaction between BHs with angular momentum or charge and their surrounding environments, leading to the extraction of energy, charge, and angular momentum from BHs. 
If there is a confinement mechanism, the scattering field will be incapable of escaping the trap and will grow over time at the expense of the BH’s energy, known as the superradiant instability or “BH bomb” \cite{Press1972}. 

In the linear regime, the superradiant instabilities of various BH-bomb systems have been studied extensively, such as BHs enclosed in a reflecting mirror \cite{Cardoso2004,Dolan2013,Degollado2013,Herdeiro2013,Hod2013a,Ferreira2017}, BHs in anti-de Sitter (AdS) backgrounds \cite{Cardoso2004a,Uchikata2009,Rosa2010,Cardoso2014,Wang2016,Ferreira2017,Uchikata2011,Gonzalez2017} and spinning BHs with massive bosonic fields in asymptotically flat spacetimes \cite{Furuhashi2004,Cardoso2005,Strafuss2005,Dolan2007,Witek2012,Dolan2013,Cardoso2018a,Dolan2018}. 
In the above models, artificial reflection boundary conditions, gravitational potential of spacetime, and self-interaction of matter fields provide confinement mechanisms respectively. 
Intriguingly, for gravitational systems that lack explicit confinement mechanisms, such as charged BHs scattered by charged scalar fields in asymptotically de Sitter (dS) spacetime, a novel instability of superradiant nature is discovered, where the superradiance is a necessary but not sufficient condition \cite{Zhu2014,Konoplya2014,Cardoso2018,Mo2018,Destounis2019,Dias2019,Liu:2020evp,Gonzalez2022,Mascher2022}.
Such an instability has also been found in charged BHs surrounded by anisotropic fluids \cite{CuadrosMelgar2021}.

To deal with saturation and the final state of superradiant instabilities, however, is beyond the scope of the linear perturbation theory and necessitates a fully nonlinear approach. 
Complicated by challenges such as very small growth rates, significant disparities in scale between the BH and superradiant fields, and inherently (3+1)-dimensional nonlinear equations \cite{Lehner2001,Choptuik2015}, only a few nonlinear evolutions of the superradiant instability have been successfully simulated in rotating BH bombs \cite{East2017,East2018,Chesler2019,Chesler2022} and charged BH bombs \cite{SanchisGual2016a,SanchisGual2016,Bosch2016,Zhang2023}.
The latter, compared with its rotating counterpart, is easier to numerically evolve to a stationary final state by reason that the charged superradiant instability has a larger growth rate and can even be explored in spherical symmetry \cite{Herdeiro2013}.
It was shown that a hairy BH with a harmonically oscillating scalar field condensate is the endpoint of charged BH bombs induced by an artificial mirror \cite{Dolan2015,SanchisGual2016a,SanchisGual2016} {or} the AdS boundary \cite{Bosch2016,Dias2017}.
 
A natural question that arises is what is the final state of a charged BH bomb in dS spacetime?
On the one hand, the results of linear perturbation theory show that such a system suffers from dynamical instability, indicating that an arbitrarily small perturbation can induce drastic changes in gravitational configuration, thereby driving the system to undergo a dynamical transition.
On the other hand, a no charged scalar hair theorem for static dS BHs has been established \cite{An2023}, which precludes the possibility of the formation of a hairy BH.
Such a contradiction motivates the implementation of nonlinear dynamics simulation.
So far, a few studies have performed nonlinear evolutions of scalar field perturbations in Reissner-Nordstr\"om (RN)-dS spacetime for strong cosmic censorship \cite{Luna2019,Zhang2019}.

To address the question, we study the nonlinear evolution of charged dS BHs scattered by a charged scalar field, based on the hyperboloidal framework, in which quasi-normal boundary conditions are inherently satisfied since no characteristics are directed to the computational domain \cite{Zenginoglu2008,Zenginoglu2011,Ripley2021}.
Hyperboloidal foliations intersect with the future cosmological horizon (respectively null infinity), where the total mass (respectively the Bondi mass) or charge of the asymptotically dS (respectively flat) spacetime can be evaluated straightforwardly \cite{Baake2016}.
At the superradiant growth stage, a rapid loss of total mass and charge of the asymptotically dS spacetime, such an unconfined system, is observed, and the initial growth rates predicted by the linear analysis are also verified.
In contrast to the charged BH bombs discussed in the previous paragraph, the scalar field cannot coexist with the central BH after saturation.
Instead, the scalar field dissipates beyond the cosmological horizon at a gradual rate, thereby reinforcing the assertion that the final state is still a bald BH with less charge \cite{An2023}. 

The paper is organized as follows: 
In Section \ref{sec2}, we introduce the hyperboloidal formalism we used. 
In Section \ref{linear}, we review the linear stability analysis of charged scalar fields for RN-dS BHs. 
In Section \ref{sec-nonlinear} we show our numerical results of non-linear evolutions for the instability with superradiant nature and discuss impacts of some free parameters. 
Finally, in Section \ref{sec_conclution} we sum up our concluding remarks.

In this paper, we use the unit $c=\epsilon_0=1$.

\section{Hyperboloidal Formalism}
\label{sec2}
In this section, we describe our hyperboloidal formalism, in which boundary conditions, field equations and foliations of spacetime are revealed.

\subsection{Hyperboloidal coordinates} \label{HPHC}
In order to achieve the spherically symmetric dynamics we are concerned with here, inspired by the hyperboloidal compactification technique \cite{Zenginoglu2008,Zenginoglu2011,Ripley2021}, we adopt a hyperboloidal coordinates with Bondi-Sachs-like gauge choices \cite{Bondi1962,Sachs1962}:
\footnote{There are some other types of hyperboloidal evolution formalism with different gauge conditions, e.g. ADM-like formulation on constant mean curvature surfaces \cite{Rinne2010,Rinne2013,Baake2016} and the Generalized BSSN or Z4 formulations \cite{VanoVinuales2015a,VanoVinuales2015,VanoVinuales2017}.}
\begin{equation} \label{ds2}
\begin{aligned}
    \mathrm{d}s^2=\frac{L^2}{Z^2}\{ 
        &  e^{-\chi}[-(1-H^2)\mathrm{d}T^2+2H\mathrm{d}T\mathrm{d}Z+\mathrm{d}Z^2] \\
        & + L^2 \mathrm{d}\Omega ^2 \},
\end{aligned}
\end{equation}
where we take the area radius $L^2/Z$ as a coordinate, $\mathrm{d}\Omega^2=\mathrm{d}\theta^2+\sin^2\theta\mathrm{d}\phi^2$ is the line element of unit sphere $S^2$, and $\chi,H$ are metric functions dependent on $T\text{ and }Z$. 
The boost function $H$ determines the locations of physical boundaries, i.e. the apparent horizon $Z_h$ and cosmological horizon $Z_c$, which are located at $H(T,Z_h)=-1 \text{ and } H(T,Z_c)=+1$ respectively. 
Additionally, the quasi-normal boundary conditions for a BH, such an open system, that the purely ingoing wave at the BH event horizon and outgoing wave at the cosmological horizon are inherently satisfied in the hyperboloidal coordinates (\ref{ds2}). 

To show this, we compute the outgoing/ingoing radial null characteristic speeds $c_{\pm}$ by solving for the eikonal equation 
\begin{equation}
    g^{\mu\nu}\xi_{\mu}\xi_{\nu}=0.
\end{equation}
Setting $\xi_{\theta}=\xi_{\phi}=0$, we obtain the outgoing radial null characteristic speed
\begin{equation}\label{outspeed}
    c_+ = \frac{\xi^+_T}{\xi^+_Z} = -H-1,
\end{equation}
and the ingoing one
\begin{equation}\label{inspeed}
    c_- = \frac{\xi^-_T}{\xi^-_Z} = -H+1.
\end{equation}
Note that at the apparent horizon $(H=-1)$ $c_+$ vanishes and $c_-$ is positive, and at the cosmological horizon $(H=1)$ $c_-$ vanishes and $c_+$ is negative, which is precisely what the quasi-normal boundary conditions require. 

\subsection{Field equations}
\label{eqs}
As a concrete example, we consider the model where a charged scalar field is minimally coupled to the Einstein-Maxwell system in asymptotically dS spacetime with the action
\footnote{Note that $A_{\mu}$ and $\psi$ here are dimensionless. 
The dimensionful physical fields are $\tilde{A_{\mu}}=A_{\mu}/\sqrt{4\pi G}, \tilde{\psi}=\psi/\sqrt{4\pi G}.$ }
\begin{equation}\label{action}
S= \frac{1}{16\pi G} \int{\mathrm{d}^4}x\sqrt{-g}\left[ R-2\Lambda -F_{\mu \nu}F^{\mu \nu}+4\mathcal{L}_m \right],
\end{equation}
\begin{equation}
    \mathcal{L}_m = -D_{\mu}\psi \overline{D^{\mu}\psi }-\mu ^2\psi \bar{\psi},
\end{equation}
where $R$ represents the Ricci scalar, $\Lambda$ represents the positive cosmological constant, $F_{\mu\nu}=\partial_\mu A_\nu-\partial_\nu A_\mu$ represents the field strength with $A_\mu$ the electromagnetic potential, $D_{\mu}=\nabla _{\mu}-iqA_{\mu}$ is the gauge covariant derivative with $q$ denoting the gauge coupling constant, and $\mu$ represents the mass of the scalar field.

Varying the action (\ref{action}), three equations of motion are obtained 
\begin{equation}
R_{\mu \nu}-\frac{1}{2}Rg_{\mu \nu}+\Lambda g_{\mu \nu}=2(T_{\mu \nu}^{\psi}+T_{\mu \nu}^{A}),
\end{equation}
\begin{equation}
\nabla _{\mu}F^{\mu \nu}=j^\nu ,
\end{equation}
\begin{equation}
\left( D_{\mu}D^{\mu}-\mu ^2 \right) \psi =0,
\end{equation}
 where the energy-momentum tensors are given by
\begin{equation}
T_{\mu \nu}^{\psi} = 
D_{\mu}\psi \overline{D_{\nu}\psi }+D_{\nu}\psi \overline{D_{\mu}\psi }-g_{\mu \nu}\mathcal{L}_m
,
\end{equation}
\begin{equation}
T_{\mu \nu}^{A}=F_{\mu \lambda}{F_{\nu}}^{\lambda}-\frac{1}{4}g_{\mu \nu}F_{\rho \sigma}F^{\rho \sigma},
\end{equation}
and the {Noether} current is given by
\begin{equation}
    j^\nu \equiv
iq\left( \bar{\psi}D^{\nu}\psi -\psi \overline{D^{\nu}\psi } \right).
\end{equation}
The model is invariant under a local $U(1)$ gauge transformation $\psi\rightarrow\psi e^{iq\alpha},A_\mu\rightarrow A_\mu+\nabla_\mu\alpha$, where $\alpha$ is a regular
real function of spacetime coordinates.

Using the hyperboloidal coordinates (\ref{ds2}), taking the gauge potential $A_\mu\mathrm{d}x^\mu = A(T,Z)\mathrm{d}T$ and introducing auxiliary variables
\begin{equation} \label{AB}
B=e^{\chi}A^\prime,
\end{equation}

\begin{equation} \label{psi_evo}
\zeta = -\frac{Z^2}{\sin\theta}\sqrt{-g}g^{T\mu}D_{\mu}\psi =\dot{\psi} -iqA\psi - H\psi^\prime ,
\end{equation}
where dot and prime denote the derivative with respect to the temporal coordinate $T$ and the radial coordinate $Z$ respectively, one can reduce the field equations to a system of equations with a simple nested structure:

{Einstein equations}
\begin{equation}\label{chi_evo}
\dot{\chi}-2H\chi ^{\prime}= 4Z\mathrm{Re}\left[ \zeta \bar{\psi}^{\prime} \right] ,
\end{equation} 

\begin{equation}\label{H_evo}
\begin{aligned}
\dot{H} - 2HH^\prime 
&=
\frac{3(1-H^2)}{Z}-Z\frac{e^{-\chi}}{L^2}+Z^3\frac{e^{-\chi}B^2}{L^2}
\\
& +\frac{e^{-\chi}L^2}{Z}\left[ \Lambda +2\mu ^2|\psi |^2 \right] ,
\end{aligned}
\end{equation} 

\begin{equation}\label{constraint_E}
    \begin{aligned}
0=   & -(1-H^2)\chi^{\prime}
    +2Z[|\zeta |^2+|\psi ^{\prime}|^2 
    +2H\mathrm{Re}\left( \zeta \bar{\psi}^{\prime} \right) ]
    \\
     & -Z^3(\frac{1-H^2}{Z^3})^{\prime}- 
    \frac{Z e^{-\chi}}{L^2}+\frac{Z^3e^{-\chi}B^2}{L^2}
     \\
     &
   +\frac{e^{-\chi}L^2}{Z}\left[ \Lambda +2\mu ^2|\psi |^2 \right],
    \end{aligned}
\end{equation}
{Maxwell equations}
\begin{equation}\label{Maxwell_T}
\dot{B} - HB^\prime= -\frac{2qL^2}{Z^2}\mathrm{Im}\left[ \bar{\psi}\psi ^{\prime}  \right] ,
\end{equation}

\begin{equation}\label{Maxwell_Z}
0=B^\prime+\frac{2qL^2}{Z^2}\mathrm{Im}\left[ \bar{\psi}\zeta \right],
\end{equation}
{and the scalar field equation}
\begin{equation}
\label{zeta_evo}
\dot{\zeta} - H \zeta^\prime=
Z^2\left[ \zeta \left( \frac{H}{Z^2} \right) ^{\prime}+\left( \frac{\psi ^{\prime}}{Z^2} \right) ^{\prime} \right]
+iqA\zeta
-\mu ^2\frac{L^2}{Z^2}e^{-\chi}\psi  .
\end{equation}

For the specific implementation of dynamical evolution, we use the so-called constrained evolution scheme.
The evolutionary equations for all the dynamical variables $\{H,\chi,\psi,\zeta,B\}$ are comprised of eqs.(\ref{psi_evo})-(\ref{H_evo}), (\ref{Maxwell_T}) and (\ref{zeta_evo}), by which we can calculate the temporal derivatives of all the dynamical variables.

Eqs.(\ref{AB}), (\ref{constraint_E}) and (\ref{Maxwell_Z}) serve as the constraint equations. 
We obtain $A$ by solving eq.(\ref{AB}) and eqs.(\ref{constraint_E}, \ref{Maxwell_Z}) are used to check the validity of our numerics (See Appendix.\ref{err} for more details).

We exclude $\dot{\chi}$ from the dynamical variables because the evolutionary equation of $\dot{\chi}$ (See \ref{dotchi}) is redundant, which can be derived from eqs.(\ref{chi_evo}), (\ref{H_evo}) and (\ref{constraint_E}) using the Bianchi identity. 
Instead, we use eq.(\ref{chi_evo}), the $Z$ component of momentum constraint, to solve $\dot{\chi}$.
The constraint equation (\ref{constraint_E}) will be preserved during the evolution, provided that it is satisfied on the initial time slice. 
See Appendix.\ref{eq_structure} for more details.


\subsection{Foliations of RN-dS} \label{RNdS_solution}
In order to simulate the dynamical process of scattering a charged BH by a scalar field, we first need to obtain the metric functions of the RN-dS BH solution in the hyperboloidal coordinates (\ref{ds2}).

Solving the Einstein-Maxwell equations with $\psi = 0$, one can find the RN-dS BH solution is described by $\chi(Z)=\chi_0=0$ and
\begin{equation} \label{H2}
    H^2(Z)=H_0^2(Z) = 1-\frac{Z^2}{L^2}+\frac{2M_0Z^3}{L^4}-\frac{Q_0^2Z^4}{L^6}+\frac{\Lambda L^2}{3},
\end{equation}
where  $M_0$ and $Q_0$ represent the total mass and charge of the BH respectively. Here are a few things to note.

First, the constant parameter $L$ cannot be arbitrarily set and it depends on $M_0,Q_0$ and $\Lambda$. 
The value of $L$ needs to be chosen such that $H_0$ is a continuous real function decreasing monotonically from 1, at the cosmological horizon $Z_c$, to $-1$, at the apparent horizon $Z_h$. 
There exists a zero $Z_0$ between $Z_c$ and $Z_h$, and by solving the set of equations
\begin{equation}\label{eq_L2&Z0}
    \begin{cases}
        H_0^2(Z_0)&=0\\
        2H_0(Z_0)H_0^\prime(Z_0)&=0
    \end{cases},
\end{equation}
one can obtain 
\footnote{In this paper, we fix the initial BH mass $M_0=0.2$ such that $L$ evaluated in (\ref{L2}) is close to 1.}
    \begin{align} 
    L^2&=\frac
    {96\gamma^3M_0^2}
    {
    3\left[\Delta^3+(\gamma-3)\Delta^2+3\gamma\right]
    -32\gamma^3M_0^2\Lambda
    }, \label{L2}
    \\
    Z_0&=(3M_0-\Delta)L^2/4Q_0^2,
    \end{align}
where $\gamma=Q_0^2/M_0^2,\ \Delta=\sqrt{9-8\gamma}$. 

Second, the function $H_0(Z)$ takes the positive square root of eq.(\ref{H2}) when $Z\in[Z_c,Z_0]$ and the negative square root when $Z\in[Z_0,Z_h]$.

\begin{figure}[htb]
    \centering
    \includegraphics[width=.7\linewidth]{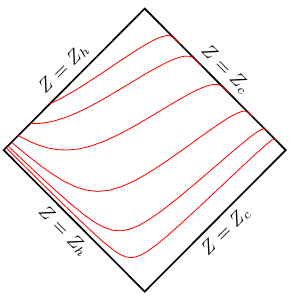}
    \caption{The Penrose diagram demonstrating the hyperboloidal foliations of RN-dS BH spacetime. The $T$-slices are the solid red lines. The hyperboloidal foliations intersect with the future cosmological horizon and penetrate the outer horizon of the RN-dS BH.}
    \label{foliation}
\end{figure}

Third, the transformation from the Schwarzschild coordinates
\begin{equation}
    \mathrm{d}s^2 = 
    -f(r)\mathrm{d}t^2+\frac{\mathrm{d}r^2}{f(r)} + r^2\mathrm{d}\Omega^2
\end{equation}
to the hyperboloidal coordinates (\ref{ds2}) can be obtained by the height function technique \cite{Zenginoglu2008,Zenginoglu2011}:
\begin{equation}
    \begin{cases}
    T& = t - h(r) \\
    Z& = {L^2}/{r}
    \end{cases}
\end{equation}
with $\mathrm{d}h/\mathrm{d}r=H_0/f$, where $h(r)$ is called the height function. 
It can be seen that our foliations penetrate both the outer horizon and the cosmological horizon (See Fig.\ref{foliation}).

\section{Linear Stability Analysis} \label{linear}
Before showing the nonlinear dynamical evolution, we would like to review the linear stability analysis of RN-dS BHs using the hyperboloidal formalism.
\footnote{The results obtained in the hyperboloidal formalism were checked with \cite{Konoplya2014,Destounis2019}.}

The linearized scalar perturbation equation on the RN-dS background is given by eqs.(\ref{psi_evo}) and (\ref{zeta_evo}), where the background electromagnetic potential is chosen as $A(Z)=-Q_0Z/L^2$. 
We follow the scheme of numerically calculating the quasi-normal modes with linearized dynamic evolution \cite{LTZZ,QNM}, which can results in a generalized eigenvalue problem \cite{Dias2010} (see also \cite{snake} for its application in inhomogeneous backgrounds and \cite{time-crystal} for that in backgrounds even without time-translation symmetry).

By expanding $\psi$ and $\zeta$ with harmonic time dependence $\psi=\tilde{\psi}(Z) e^{-i\omega T},\zeta=\tilde{\zeta}(Z) e^{-i\omega T}$, the temporal derivatives are replaced by $-i\omega$.
Meanwhile, the $Z$ coordinate is discretized with Chebyshev-Gauss-Lobatto grid points $\{Z_i\}$ and then the radial derivatives can be replaced by the corresponding differentiation matrix $\bm{D}$ \cite{Trefethen2000,Press2007}.
Thus, the quasi-normal frequencies $\omega$ can be obtained by solving the resulting generalized eigenvalue equation
\footnote{One has to be careful about false modes caused by the discretization, which can be examined by varying the grid number.}
\begin{equation} \label{gener-eig}
\left( \bm{A}-\omega \bm{B} \right) \left( \begin{array}{c}
	\tilde{\zeta}\\
	\tilde{\psi}\\
\end{array} \right) =0,
\end{equation}
with
\begin{equation}
    \bm{A} = 
    \begin{pmatrix}
        \bm{1} &
\bm{H_0}\bm{D}-iqQ_0\frac{\bm{Z}}{L^2}
 \\
\bm{D}\bm{H_0}-\frac{2}{\bm{Z}}\bm{H_0}-iqQ_0\frac{\bm{Z}}{L^2} &
\bm{D}^2-\frac{2}{\bm{Z}}\bm{D}-\mu ^2\frac{L^2}{\bm{Z}^2}
    \end{pmatrix},
\end{equation}

\begin{equation}
    \bm{B} = -i 
    \begin{pmatrix}
    \bm{0} & \bm{1}
    \\
    \bm{1} & \bm{0}
    \end{pmatrix},
\end{equation}
where $\bm{1},\bm{0}$ represent the identity matrix and the null matrix respectively, and $\bm{Z} ,\ \bm{H_0}$ represent diagonal matrices of the grid points $\{Z_i\}$ and the grid function $\{H_0(Z_i)\}$ respectively.
As mentioned in Sec.\ref{HPHC}, the quasi-normal boundary conditions are inherently satisfied in the hyperboloidal coordinates.

As already found in \cite{Zhu2014,Konoplya2014,Destounis2019}, the occurrence of instability in this system mainly depends on the charge coupling $qQ_0$, the scalar mass $\mu$ and the cosmological constant $\Lambda$. 
It was shown that for small enough $\mu$ (less than a critical mass $\mu_c$), there exists an interval $\lambda_{\text{min}}<qQ_0<\lambda_{\text{max}}$ where the unstable superradiant modes could occur
\footnote{We only consider the $qQ>0$ case due to the symmetry of eq.(\ref{gener-eig}) $qQ_0\rightarrow -qQ_0, \text{Re}[\omega]\rightarrow-\text{Re}[\omega]$.}. 

\begin{figure}[htb]
    \centering
    \includegraphics[width=\linewidth]{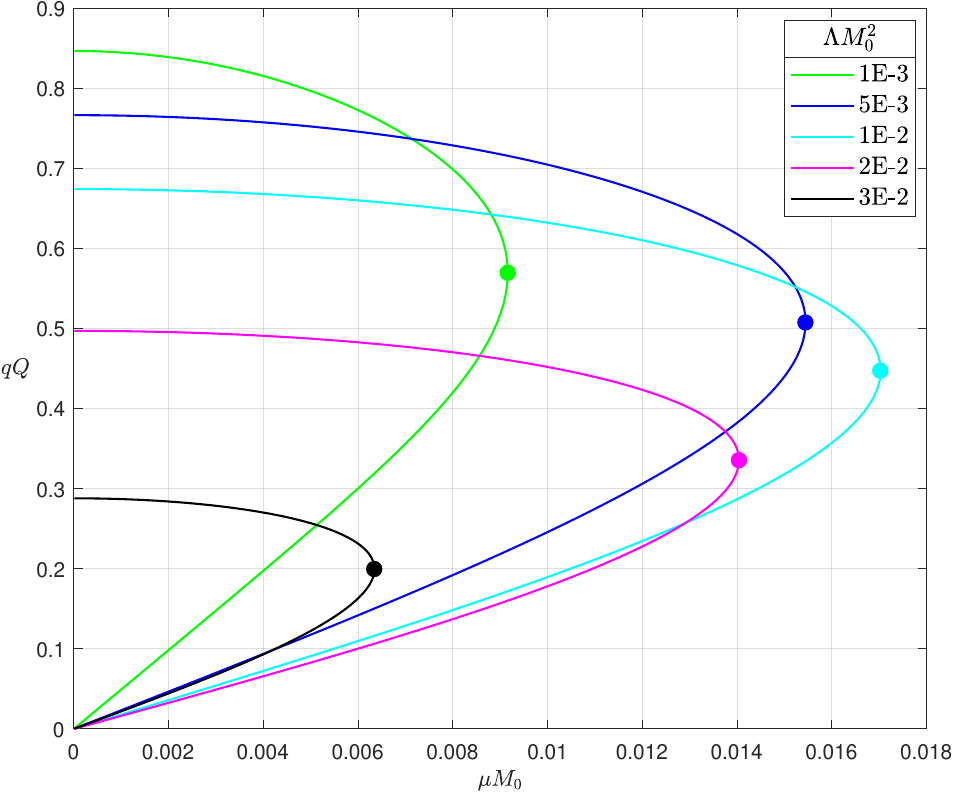}
    \caption{The unstable charge coupling interval $(\lambda_{\text{min}},\lambda_{\text{max}})$ as a function of the scalar mass $\mu$ when $M_0=0.2, Q_0=0.3M_0$. 
    Curves of different colors represent different values of the cosmological constant $\Lambda$.
    Filled dots correspond to the critical scalar mass $\mu_c$. 
    Lines above the filled dots represent $\lambda_{\text{max}}(\mu)$ and lines below the filled dots represent $\lambda_{\text{min}}(\mu)$.}
    \label{lamb_mu_La}
\end{figure}

As shown in Fig.\ref{lamb_mu_La}, while $\mu$ decreases from $\mu_c$ to zero, $\lambda_{\text{max}}$ increases, whereas $\lambda_{\text{min}}$ decreases to zero (Also see Fig.5 in \cite{Destounis2019}).
In addition, the critical mass $\mu_c$ increases first and then decreases with the increment of $\Lambda$. While $\Lambda$ reaches beyond a critical value $\Lambda_c$, no instability is observed. 

\begin{figure}[htb]
    \centering
    \includegraphics[width=\linewidth]{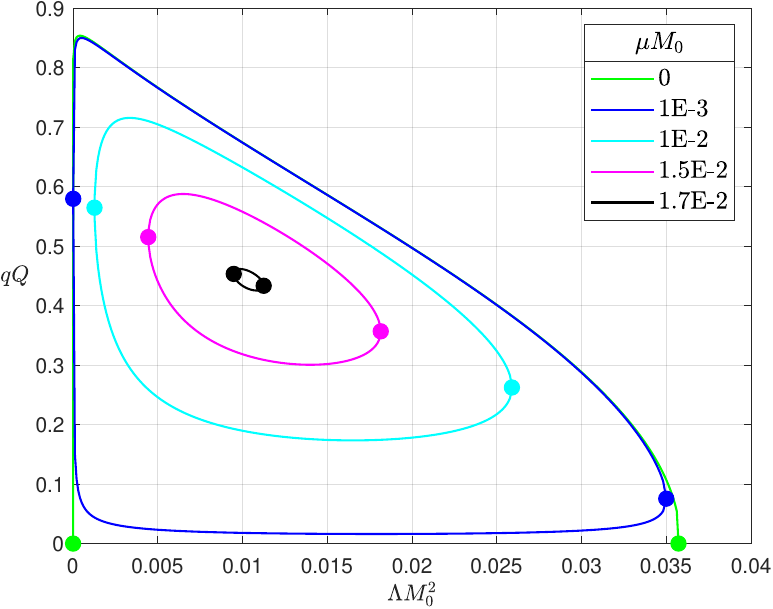}
        \caption{The unstable charge coupling interval $(\lambda_{\text{min}},\lambda_{\text{max}})$ as a function of the cosmological constant $\Lambda$ when $M_0=0.2, Q_0=0.3M_0$. 
        Curves of different colors represent different values of the scalar mass $\mu$.
        Filled dots correspond to the critical cosmological constant $\Lambda_c^\prime$ (left) and $\Lambda_c$ (right). 
        Lines above the filled dots represent $\lambda_{\text{max}}(\Lambda)$ and lines below the filled dots represent $\lambda_{\text{min}}(\Lambda)$. 
        For $\mu=0$, $\lambda_{\text{min}}(\Lambda)\equiv0$ and $\Lambda_c^\prime=0.$}
    \label{lamb_La_mu}
\end{figure}

For the case of a massless scalar field, unstable modes appear for arbitrarily small $\Lambda$ \cite{Konoplya2014,Destounis2019}. 
However, we find that the situation with massive scalar fields is completely different.
As shown in Fig.\ref{lamb_La_mu}, when $\mu>0$, there exists a lower bound $\Lambda_{c}^\prime>0$, below which no instability can be observed.
As the scalar mass $\mu$ increases to a critical value, the region $(\Lambda_{c}^\prime,\Lambda_{c})$ shrinks until it disappears.

\section{Nonlinear Dynamical Evolution} 
\label{sec-nonlinear}
In this section, we present our numerical evolutions of the Einstein-Maxwell-Scalar-$\Lambda$ (EMS$\Lambda$) system derived in the Sec.\ref{eqs}.
In Sec.\ref{numerical setup} we briefly describe our numerical setup. 
The initial data for RN-dS BHs with charged scalar fields are constructed in Sec.\ref{init}.  
In Sec.\ref{General picture}, we will describe a general picture for the full nonlinear evolution of the instability driven by superradiance.
Finally in Sec.\ref{Detailed description}, we discuss the effects of four free parameters on the dynamics of the system: the scalar mass $\mu$, the cosmological constant $\Lambda$, the charge coupling $qQ_0$ \footnote{We only show the result for $Q_0=0.3M_0$, for results for other value of $Q_0$ are qualitatively the same.} and the scalar field amplitude $k$.

\subsection{Numerical setup} \label{numerical setup}
To efficiently perform longtime evolutions, we employ explicit fourth-fifth order Runge-Kutta method \cite{Dormand1980,Shampine1997} along the time $T$ direction and Chebyshev pseudo-spectral method (see e.g.\cite{Trefethen2000,Press2007}) in the radial $Z$ direction.

The simulation takes place within a larger domain $Z\in[Z_1,Z_2]$ than the region $[Z_c,Z_h]$ between the apparent horizon $Z_h$ and cosmological horizon $Z_c$, which ensures that the physical evolution in $[Z_c,Z_h]$ can be fully determined by initial data. 
Note that $\dot{H}=-2Z_c|\zeta+\bar{\psi}^\prime|^2\leq0$ and $H^\prime<0$ at $Z_c$, thereby $\dot{Z}_c=-\dot{H}/H^{\prime}|_{Z_c}\leq0$, which means the radius of cosmological horizon does not decrease during evolution.
Thus, $Z_1$ needs to be small enough to prevent the cosmological horizon $Z_c$ from expanding beyond $Z_1$ during evolution. We empirically set $Z_1 = 0.9Z_c$ and $Z_2 = 1.01Z_h$ in our simulations.

As discussed in Sec.\ref{HPHC}, we do not need to impose physical boundary conditions\footnote{A boundary condition for the gauge field $A$ is still needed. We set $A(T,Z_1)=0$ in our simulations.}, even if our actual simulation domain $[Z_1,Z_2]\supset [Z_c,Z_h]$.

\subsection{Initial data} \label{init}
As we have five first-order temporal derivative equations (\ref{psi_evo})-(\ref{H_evo}), (\ref{Maxwell_T}), (\ref{zeta_evo}) and two constraints eqs.(\ref{constraint_E}), (\ref{Maxwell_Z}), {the initial value can be completely determined by three of the five dynamical variables} $\{H,\chi,\psi,\zeta,B\}$. 
In our case, we freely choose the initial data of $\{\psi,\zeta,\chi\}$ and solve $\{H,B\}$ via the two constraints.

We impose the following initial scalar perturbation on the seed RN-dS BH:
\begin{equation} \label{psi_init}
    \psi(T=0,Z) = k \exp[{-(\frac{Z-c}{w})^2}],
\end{equation}
\begin{equation} \label{zeta_init}
    \zeta(T=0,Z) = -H\psi^\prime,
\end{equation}
where $k$ is the amplitude, $c$ is the centre, and $w$ is the width of initial Gaussian wave packet. 
{For such a form of perturbation, on the one hand, }the total mass of the spacetime, evaluated by the rescaled Misner-Sharp mass (See \ref{mass_MS}) at the cosmological horizon $Z_c$, increases with the amplitude $k$. 
{On the other hand, the total charge of the perturbed system remains unchanged, indicating the initial scalar perturbation is neutral. Such a result can be derived from the constraint eq.(\ref{Maxwell_Z}), where $B(T=0,Z)$ is a constant associated with the BH initial charge $Q_0$ (See \ref{BH-charge}).}
We have also used other functional forms of perturbation, such as linear functions. All the numerical results are qualitatively the same.

The existence of the scalar perturbation $\psi$ will deform the RN-dS metric functions $\chi_0(Z)$ and $H_0(Z)$. We choose $\chi(T=0,Z)$ to satisfy $\dot{\chi}=0$, which gives
\begin{equation} \label{chi_init}
\chi ^{\prime}=2Z|\psi ^{\prime}|^2.
\end{equation}
This choice makes the constraint equation (\ref{constraint_E}) equivalent to $\dot{H}=0$, that is,
\begin{equation} \label{H_init}
\left( \frac{1-H^2}{Z^3} \right) ^{\prime}=\frac{e^{-\chi}}{L^2}\left[ -\frac{1}{Z^2}+\frac{L^4}{Z^4}\left( \Lambda +2\mu ^2|\psi |^2 \right) +B^2 \right].
\end{equation}
Integrating eq.(\ref{H_init}) numerically to solve $H(T=0,Z)$ may require a little trick, see Appendix.\ref{construction-H_init} for more details.

\subsection{General picture}\label{General picture}
{In order to give a general picture of the evolution of the EMS$\Lambda$ system, we show the characteristic behaviors of some key physical quantities including} the scalar field value $\psi_h$ on the apparent horizon, the scalar field energy $E_\psi$ and charge $Q_\psi$, as well as the BH charge $Q_h$, the BH Mass $M_\text{B}$, irreducible mass $M_{\text{irr}}$, and the total charge $Q_{\text{tot}}$ and total mass $M_{\text{MS}}$ of the spacetime (See Appendix.\ref{Physical quantities} for detailed calculations). 

\begin{figure*}
    \includegraphics[width=8.2cm]{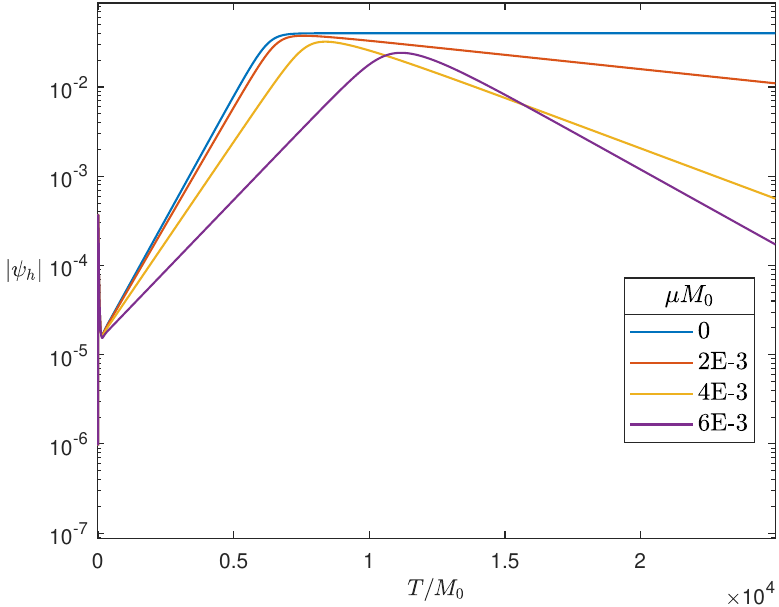}
    \includegraphics[width=8.4cm]{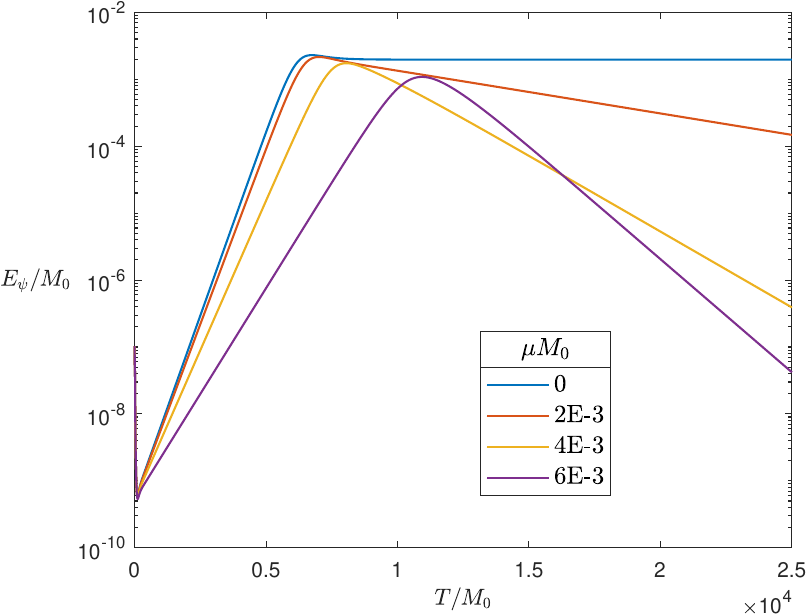}
    \caption{Time evolution of the absolute scalar field value $|\psi_h|$ on the apparent horizon (left panel) and the scalar field energy $E_\psi$ (right panel) with $M_0=0.2, Q_0=0.3M_0, \Lambda M_0^2=10^{-3}, qQ_0=0.55$. 
    Curves of different colors represent different values of the scalar mass $\mu$. }
    \label{psi_fig}
\end{figure*}

The time evolution of $\psi_h$ exhibited in the left panel of Fig.\ref{psi_fig} shows two distinct stages. During the first stage, the so-called \textit{superradiant growth stage}, the scalar field grows exponentially, as expected by the linear {stability} analysis (see Table.\ref{tab_mu}). 
During the second stage, the \textit{relaxation stage}, the scalar field reaches saturation and then decays slowly (exponentially as well).
Since the spacetime background changes also slowly at the \textit{relaxation stage}, we can adopt adiabatic approximation to analyze the decay rates $1/\tau_r$ of the relaxing scalar fields at a linear level \cite{Brito2015} (see also Table.\ref{tab_mu}).
The similar behavior of the corresponding evolution of the scalar field energy $E_\psi$ is exhibited in the right panel of Fig.\ref{psi_fig}.

\begin{figure*}
    \centering
    \includegraphics[width=8.3cm]{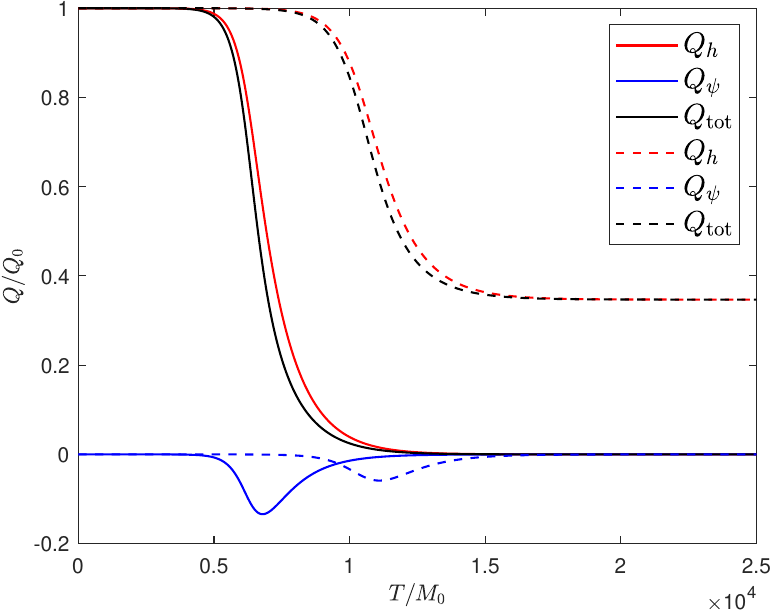}
    \includegraphics[width=8.4cm]{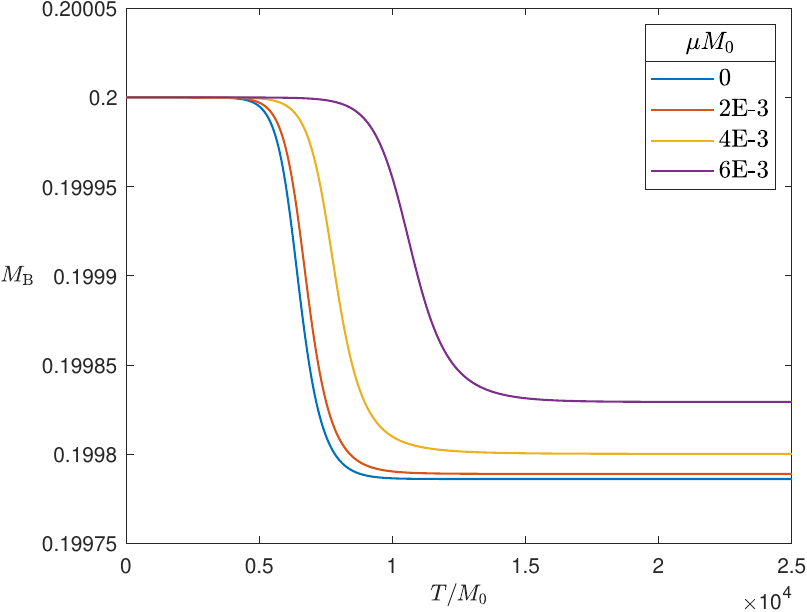}
    \caption{
    Left panel:
    Time evolution of the BH charge $Q_h$, the scalar field charge $Q_\psi$ and the total charge $Q_{\text{tot}}$ with $M_0=0.2, Q_0=0.3M_0, \Lambda M_0^2=10^{-3}, qQ_0=0.55$, $\mu M_0=0$ (solid line) and $6\times10^{-3}$ (dotted line). 
    Accompanied by the exponential growth of the scalar field, the charge of the BH is extracted drastically, and escapes beyond the cosmological horizon.
    Right panel:
    Time evolution of the BH mass $M_{\text{B}}$ with $M_0=0.2, Q_0=0.3M_0, \Lambda M_0^2=10^{-3}, qQ_0=0.55$. 
    Curves of different colors represent different values of the scalar mass $\mu$.
    }
    \label{Q}
\end{figure*}

As depicted in the left panel of Fig.\ref{Q}, at the \textit{superradiant growth stage}, a large portion of the scalar field with {the opposite sign to the BH charge} is absorbed into the apparent horizon, leading to a sharp reduction in the BH charge.
Meanwhile, {most} of the scalar field with {the same sign as the BH charge} is repelled beyond the cosmological horizon, resulting in a greater decrease in the system's total charge and a negative net charge of the scalar field. 
Additionally, the BH experiences a slight reduction in mass during the \textit{superradiant growth stage} (See the right panel of Fig.\ref{Q}).

\begin{figure*}
    \includegraphics[width=8.3cm]{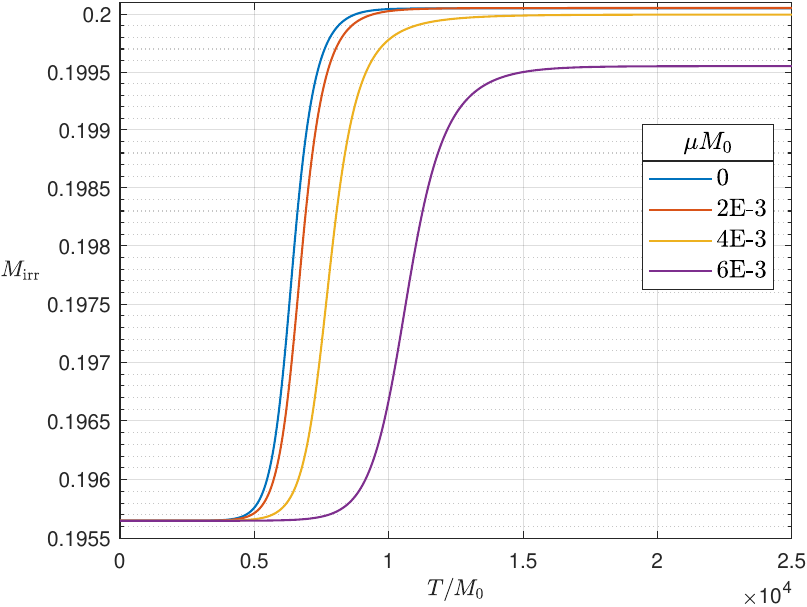}
    \includegraphics[width=8.3cm]{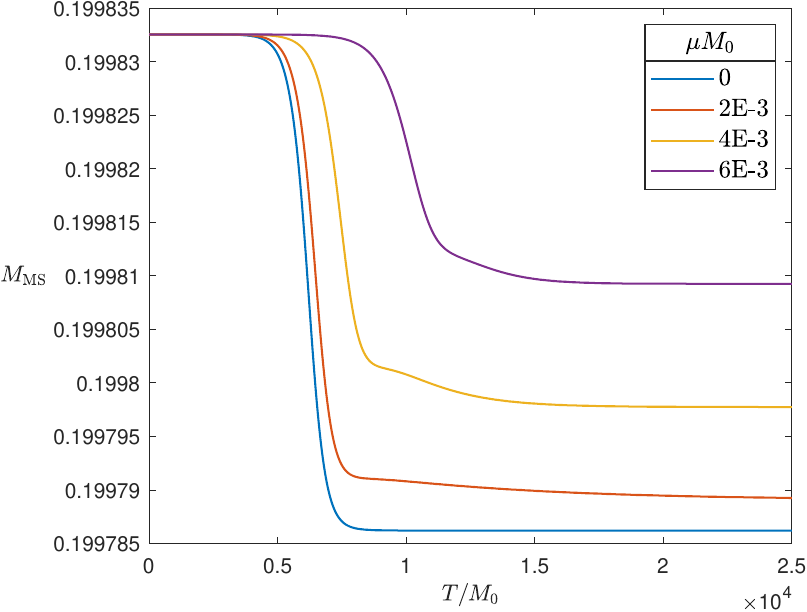}
    \caption{Time evolution of the irreducible mass $M_{\text{irr}}$ (left panel) and the total mass of the spacetime $M_{\text{MS}}$ (right panel) with $M_0=0.2, Q_0=0.3M_0, \Lambda M_0^2=10^{-3}, qQ_0=0.55$. Curves of different colors represent different values of the scalar mass $\mu$.}
    \label{Mass}
\end{figure*}

As shown in Fig.\ref{Mass}, the irreducible mass $M_{\text{irr}}$ increases rapidly at the \textit{superradiant growth stage}, then remains nearly a constant at the \textit{relaxation stage}.
{Such a result that the irreducible mass does not decrease throughout the evolution abides by the area law.}
Finally, shown in the right panel of Fig.\ref{Mass}, the total mass of spacetime does not increase throughout the evolution, as a result of charged scalar radiation.   

\begin{table*}
    \centering
    \caption{Summary of physical quantities for the runs with $M_0=0.2, Q_0=0.3M_0, \Lambda M_0^2=10^{-3}, qQ_0=0.55,k=10^{-4}$, and different values of the scalar mass $\mu$. 
    We have checked that the e-folding time obtained as the best fit of the form $|\psi|\sim e^{T/\tau_s}$ during the \textit{superradiant growth stage} matches the prediction of the imaginary part of the unstable mode $\omega_I$ in the linearized case.
    $Q_h^{\text{f}},Q_\psi^{\text{f}},Q_{\text{tot}}^{\text{f}}$ and $M_{\text{B}}^{\text{f}}$ are evaluated at $T=2.5\times10^4M_0$, and we have checked $Q_h+Q_\psi=Q_{\text{tot}}$ during the evolution.
     We have also checked that the e-folding time obtained as the best fit of the form $|\psi|\sim e^{T/\tau_r}$ during the \textit{relaxation stage} approximates to the prediction of the imaginary part $\omega_I^\prime$ of the dominant mode calculated by $Q_h^{\text{f}},M_{\text{B}}^{\text{f}}$.
     }
    \begin{tabular}{|c|cc|cc|ccc|c|cc}
         \hline
         $\mu M_0$ &   $M_0/\tau_s$    &    $\omega_IM_0$  
         &  $\lambda_{\text{min}}$ & $\lambda_{\text{max}}$ 
         & $Q_h^{\text{f}}/Q_0$ & $Q_\psi^{\text{f}}/Q_0$ & $Q_{\text{tot}}^{\text{f}}/Q_0$
         & $M_{\text{B}}^{\text{f}}/M_0$
         & $M_{\text{B}}^{\text{f}}/\tau_r$    &    $\omega_I^\prime M_{\text{B}}^{\text{f}}$
        \\ \hline
        0   & 0.00126749    & 0.00126817 
        &0      & 0.8466    & 1.33E-7   &-4.9E-8     & 8.4E-8
        & 99.89\%
        & 0 & 0
        \\ \hline
        2E-4& 0.00126690    & 0.00126758   
        &0.0098 &  0.8466   & 1.62E-7   &-5.8E-8     & 1.04E-7
        & 99.89\%
        & -7.36E-7 & -7.15E-7
        \\\hline
        2E-3& 0.00120829    & 0.00120865 
        &0.0984 & 0.8390    & 0.0024    &-8.3E-5     & 0.0023
        & 99.89\%
        & -7.36E-5 & -7.16E-6
        \\ \hline
        4E-3& 0.00102937    & 0.00102935  
         &0.1975 & 0.8153   & 0.1232    &-1.1E-5     & 0.1232
        & 99.90\%
        & -2.61E-4 & -2.53E-4
         \\ \hline
        6E-3& 0.00072810    & 0.00072797
        &0.3004 & 0.7728    & 0.3470    &-2.3E-6     & 0.3470
        & 99.91\%
        & -3.89E-4 & -3.79E-4
         \\ \hline
    \end{tabular}
    \label{tab_mu}
\end{table*}

\subsection{Detailed description}\label{Detailed description}

In the previous subsection, our results have shown that {the scattering process of charged scalar fields can extract charge and mass from the central RN-dS BHs through superradiance.}
Since the occurrence of this instability in this system is strongly linked to the charge coupling, we focus on the effects of the four free parameters on the amount of charge extracted from the BH in this subsection.

Predictably, $\lambda_{\text{min}}$ denotes the saturation point of the instability.
We find that in the case of the massless scalar field, where $\lambda_{\text{min}}=0$, the BH charge decays exponentially.
However, for massive scalar fields ($\lambda_{\text{min}}>0$), the amount of charge extracted from the BH is associated with the difference $qQ_0-\lambda_{\text{min}}$. 
More specifically, the bigger the difference $qQ_0-\lambda_{\text{min}}$, the more charge of the BH is extracted. 
In the following, we run simulations with different parameters to verify this claim.

\subsubsection{Impact of the scalar mass and cosmological constant}
As discussed in Sec.{\ref{linear}}, {the lower bound of the charge coupling interval governing the instability of the system}, denoted as $\lambda_{\text{min}}$, is subject to influence from the scalar mass $\mu$ and the cosmological constant $\Lambda$. 
Thus, we {adjust} $\mu$ {and} $\Lambda$ to vary $\lambda_{\text{min}}$ {while fixing} the initial charge coupling $qQ_0$.

\begin{figure}[htb]
    \centering
    \includegraphics[width=\linewidth]{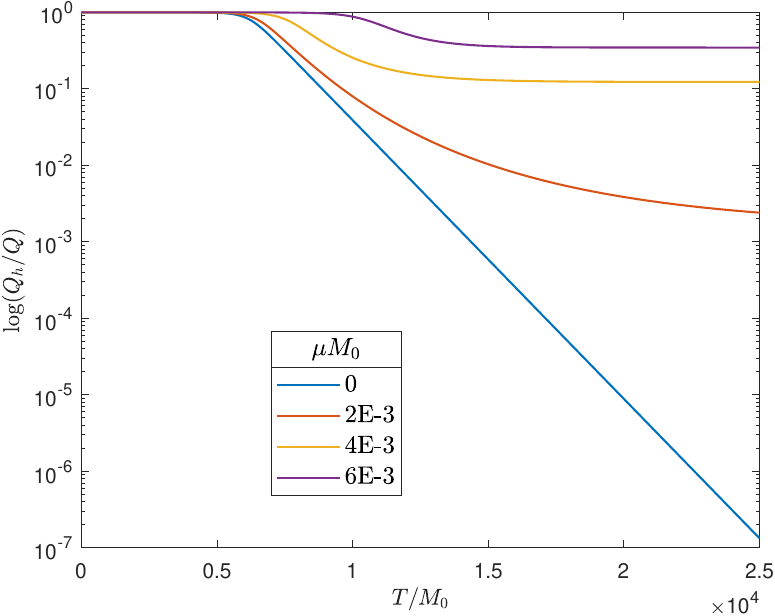}
    \caption{Time evolution of the BH charge $Q_h$ with $M_0=0.2, Q_0=0.3M_0, \Lambda M_0^2=10^{-3}, qQ_0=0.55$. 
    Curves of different colors represent different values of the scalar mass $\mu$.
    Note that the ordinate is logarithmic.}
    \label{ln_Q_t_mu}
\end{figure}

As shown in Fig.\ref{ln_Q_t_mu} and Table.\ref{tab_mu}, the more massive the scalar field is, the less charge the BH loses. 
It is noteworthy that in the case of the massless scalar field, the charge of the BH decays exponentially, which implies the final BH could be almost neutral. 

\begin{figure*}
    \centering
    \includegraphics[width=8.3cm]{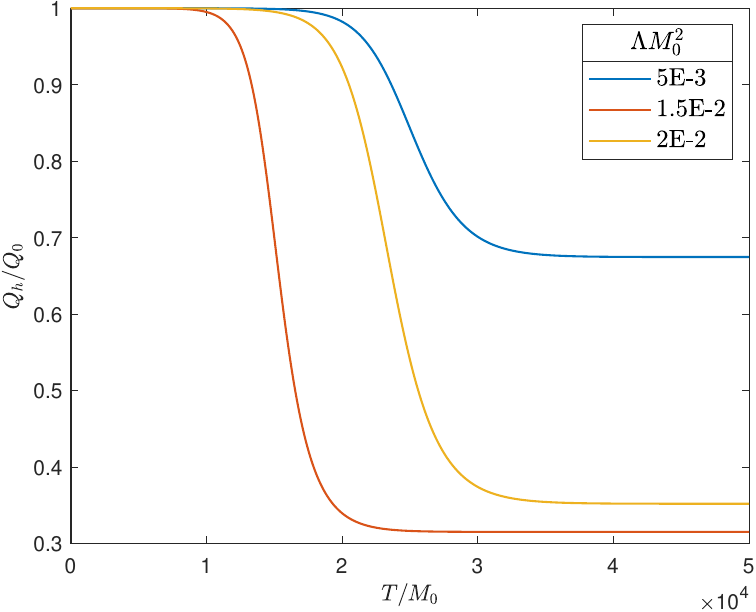}
    \includegraphics[width=8.3cm]{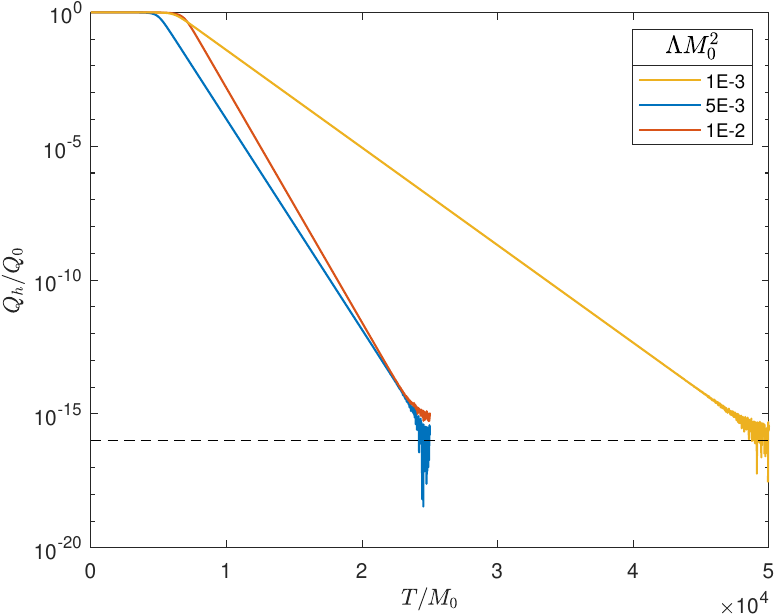}
    \caption{Time evolution of the BH charge $Q_h$ with $\mu M_0=10^{-2}$ (left panel) and $\mu M_0=0$ (right panel), when $M_0=0.2, Q_0=0.3M_0, qQ_0=0.3$ (left panel) and $qQ_0=0.55$ (right panel).
    Curves of different colors represent different values of the cosmological constant $\Lambda$.
    Note that the ordinate of the right panel is logarithmic.
    The horizontal dashed line line in the right panel represents the nominal round-off level 1E-16 for the double precision employed.}
    \label{Q_t_La}
\end{figure*}

Results with different $\Lambda$ are shown in Fig.\ref{Q_t_La} and Table.\ref{tab_La}.
In the context of the massive scalar field, the RN-dS BH with the moderate $\Lambda$, thereby smaller $\lambda_{\text{min}}$, loses more charge. 
However, in the case of the massless scalar field, $\lambda_{\text{min}}$ always vanishes for $\forall\Lambda\in(0,\Lambda_c)$.
Consequently, for different $\Lambda$, the numerical values of the BH charge all decrease exponentially to $10^{-16}$, the nominal round-off level for the double precision employed.

\begin{table}
    \centering
    \caption{Summary of physical quantities for the runs with different values of the cosmological constant $\Lambda$, when $M_0=0.2, Q_0=0.3M_0, \mu M_0=0.01, qQ_0=0.3$ and $k=10^{-4}$.
    $Q_h^{\text{f}},Q_\psi^{\text{f}},Q_{\text{tot}}^{\text{f}}$ and $M_{\text{B}}^{\text{f}}$ are evaluated at $T=5\times10^4M_0$.} 
    \begin{tabular}{|c|cc|ccc|c}
        \hline
        $\Lambda M_0^2$  &   $\lambda_{\text{min}}$ & $\lambda_{\text{max}}$
        & $Q_h^{\text{f}}/Q_0$ & $Q_\psi^{\text{f}}/Q_0$ & $Q_{\text{tot}}^{\text{f}}/Q_0$
        & $M_{\text{B}}^{\text{f}}/M_0$
        \\
        \hline
        0.5E-2  &0.2458 & 0.7046 
        & 67.47\% & 1.4E-7 & 67.47\%
        & 99.88\%
        \\ 
        \hline
        1.5E-2  &0.1742 & 0.5471   
        & 31.54\% & 1.6E-11 & 31.54\%
        & 99.64\%
        \\ 
         \hline
        2.5E-2  &0.2177 & 0.3231   
        & 35.21\% & 5.7E-8 & 35.21\%
        & 99.58\%
        \\
        \hline
    \end{tabular}
    \label{tab_La}
\end{table}

Thus, we can conclude that when the parameters except $\mu$ and $\Lambda$ are fixed, the smaller $\lambda_{\text{min}}$ is, the more complete {the} charge extraction will be. 

\subsubsection{Impact of the charge coupling}
\begin{figure}[htb]
    \centering
    \includegraphics[width=.9\linewidth]{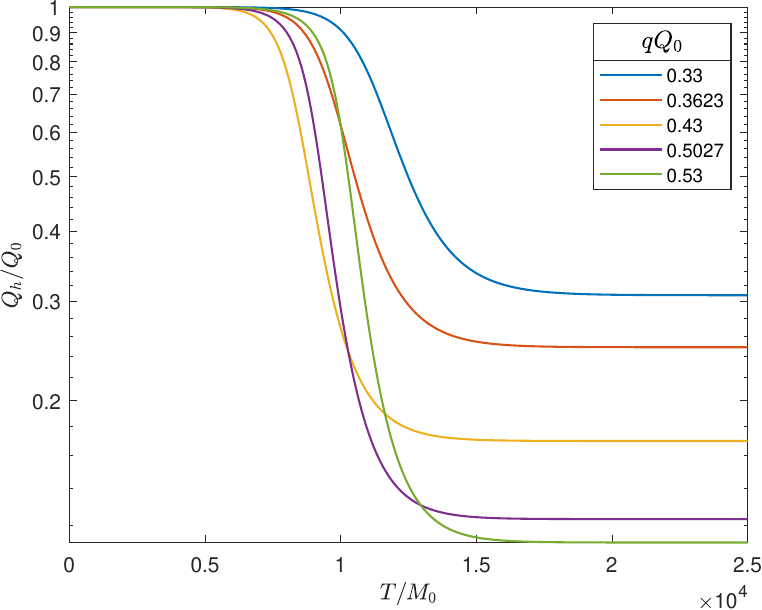}
    \caption{Time evolution of BH charge $Q_h$ with $M_0=0.2, Q_0=0.3M_0, \Lambda M_0^2=10^{-2}, \mu M_0=10^{-2}, k=10^{-4}$. 
    Curves of different colors represent different values of the initial charge coupling $qQ_0$.
    The unstable charge coupling interval  $(\lambda_{\text{min}},\lambda_{\text{max}})$ is $(0.1896, 0.6322)$ in this case. 
    Note that the ordinate is logarithmic.}
    \label{impact_qQ}
\end{figure}

To investigate the impact of the charge coupling, we {keep} the $\mu$ and $\Lambda$ to fix $\lambda_{\text{min}}$, and then vary the initial $qQ_0$. 
As shown in Fig.\ref{impact_qQ}, when $\lambda_{\text{min}}$ is fixed, the larger initial charge coupling $qQ_0$ (not larger than $\lambda_{\text{max}}$), the more charge is extracted from the BH.

Thus, taking the result of the previous subsubsection into consideration, we conclude that the amount of charge extracted from the BH is associated with the difference $qQ_0-\lambda_{\text{min}}$. 

\subsubsection{Impact of the scalar field amplitude}
\begin{figure}[htb]
    \centering
    \includegraphics[width=.9\linewidth]{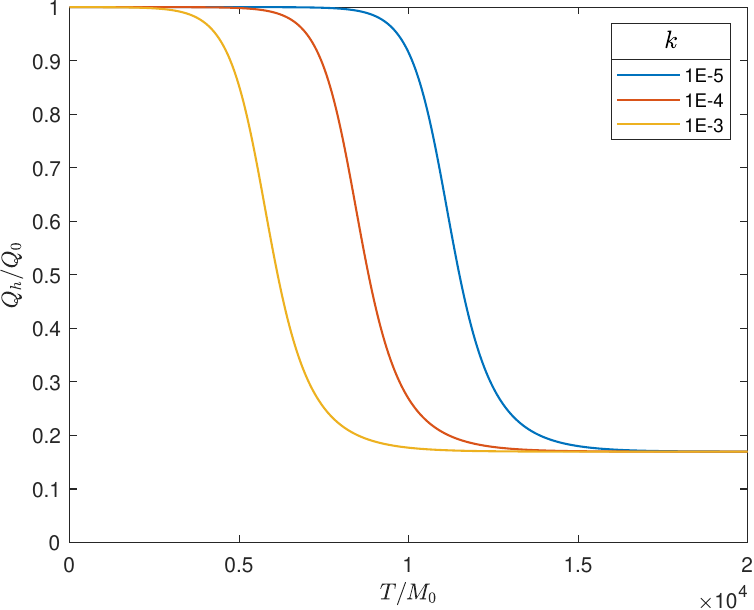}
    \caption{Time evolution of BH charge $Q_h$ with $M_0=0.2, Q_0=0.3M_0, \Lambda M_0^2=10^{-2}, \mu M_0=10^{-2}, qQ_0=0.43$.
    Curves of different colors represent different values of the scalar field amplitude $k$.}
    \label{impact_k1}
\end{figure}

One might also think that the amount of charge extracted from the BH could be affected by the scalar field amplitude $k$. 
However, we find that the scalar field amplitude $k$ hardly affects the amount of charge extracted from the BH, and that a larger $k$ could advance the start time of the \textit{superradiant growth stage} (See Fig.\ref{impact_k1}).

\begin{figure*}
    \centering
    \includegraphics[width=8.5cm]{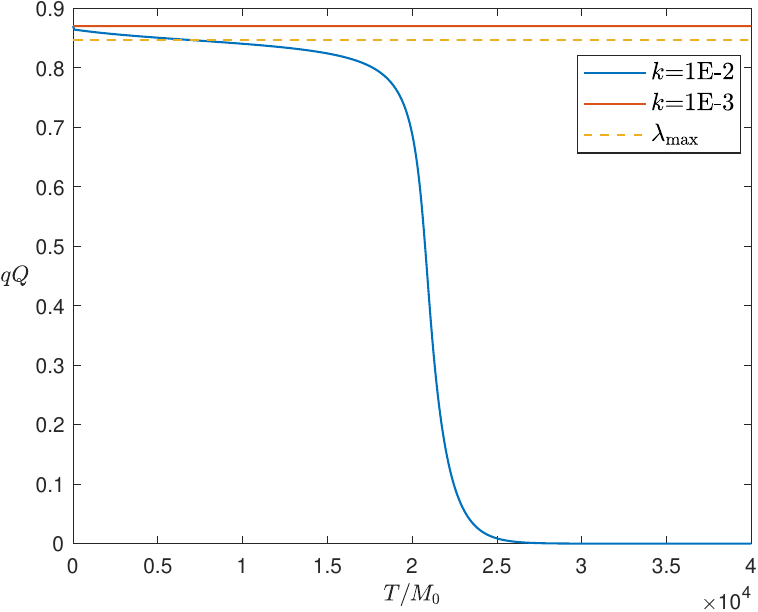}
    \includegraphics[width=8.1cm]{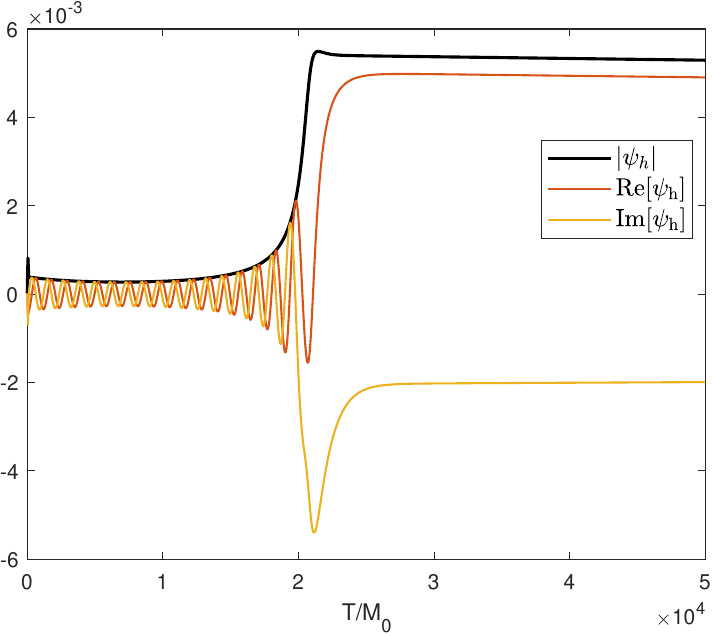}
    \caption{
    Left panel: Time evolution of charge coupling $qQ$ in the case of $M_0=0.2, Q_0=0.3M_0, \Lambda M_0^2=10^{-3}, \mu M_0=2\times10^{-4}, qQ_0=0.87$, where $\lambda_{\text{max}}=0.8466$ is represented by a dashed line.
    The red line represents the small perturbation with $k=10^{-3}$, and the blue line represents the significant perturbation with $k=10^{-2}$.
    Right panel: Time evolution of the absolute value $|\psi_h|$ (black), the real part $\text{Re}[\psi_h]$ (red) and the imaginary part $\text{Im}[\psi_h]$ (orange) of the scalar field value on the apparent horizon, corresponding to the case of $k=10^{-2}$ in the left panel.}
    \label{impact_k2}
\end{figure*}

More intriguingly, a significant perturbation with an initial charge coupling $qQ_0$ beyond $\lambda_{\text{max}}$ can also trigger this instability (See Fig.\ref{impact_k2}).
In the case of the right panel of Fig.\ref{impact_k2}, the superradiant mode is stable at the early stage.
Nevertheless, a larger initial amplitude $k$ enhance the early charge extraction.
Thus, an initial charge coupling $qQ_0$ beyond $\lambda_{\text{max}}$ can enter the unstable interval $\left(\lambda_{\text{min}},\lambda_{\text{max}}\right)$ at some point, where the superradiant mode becomes unstable.

\section{Conclusions} \label{sec_conclution}
In the present work, we study the dynamical process of charged {dS BHs} scattered by a charged scalar field with the hyperboloidal formalism. 

{For studying gravitational systems in asymptotically dS spacetime,} the hyperboloidal formalism has many advantages.
{On the one hand, the structure of field equations is simple.
On the other hand, since the foliations penetrate both the apparent horizon and the cosmological horizon, the physical boundary conditions are inherently satisfied, which greatly simplifies the work of stability analysis and dynamical simulation of the system.}
In addition, the hyperboloidal formalism we employed is also well-suited for gravitational systems in asymptotically flat spacetime without recourse to the conformal transformation.

The linear stability analysis of the RN-dS BH scattered by a charged scalar field is reviewed within the hyperboloidal framework.
Our results agree well with those in the previous literature.
Moreover, we find that black holes of intermediate size, relative to the radius of the cosmological horizon, are prone to instabilities induced by massive scalar fields.

The nonlinear numerical simulations are implemented to reveal the real-time dynamical process of such an instability.
Our main result is that the scalar field grows exponentially at the \textit{superradiant growth stage}, {leading to a substantial loss of the charge of the black hole.} The scalar field then slowly dissipates beyond the cosmological horizon, {causing the total charge of the spacetime to decrease and eventually} leaving behind a bald black hole.
{The} massless scalar field relaxes to a constant value at late-time and the charge of the BH decreases exponentially, which implies the final BH could be almost neutral.
However, for massive scalar fields, the amount of charge extracted from the BH is associated with the difference $qQ_0-\lambda_{\text{min}}$. 
In addition, significant perturbations with an initial charge coupling $qQ_0$ beyond $\lambda_{\text{max}}$ can also trigger such an instability. 

{As is known to all, in addition to charged BHs, rotating BHs can also trigger superradiance.
Therefore, a further meaningful study is the superradiance process of Kerr BHs in asymptotically dS spacetime scattered by a scalar field \cite{Tachizawa1993,Anninos2010,Georgescu2014,Zhang2014,Bhattacharya2018}.
Such a process will induce the release of gravitational waves, which is of great astronomical observation significance.
Furthermore, there are indications that critical phenomena {between the more/less charged BH transition (analogous to the gravitational collapse \cite{Choptuik1993,Abrahams1993,Evans1994,Gundlach1995,Koike1995,Garfinkle1998,Choptuik2004,Gundlach2007} or the bald/scalarized BH transition \cite{Zhang2022,Zhang2022a,Jiang2023,Chen2023,Liu2023})} may exist in the superradiance process of the EMS$\Lambda$ system, which we reserve for further study.}

\begin{acknowledgments}
We would like to thank Ulrich Sperhake and Carlos A.R. Herdeiro for their helpful discussions. Zhen-Tao He would like to thank Jia Du for his help with Mathematica, Zhuan Ning and Yu-Kun Yan for discussions about the generalized eigenvalue method, and Meng Gao for discussions about Newton-Raphson iteration algorithm. This work is partly supported by the National Key Research and Development Program of China (Grant No.2021YFC2203001). This work is supported in part by the National Natural Science Foundation of China under Grants No. 12035016, No. 12075026, No. 12275350, No. 12375048, No. 12375058 and. No. 12361141825.
\end{acknowledgments}
\appendix
\section{Calculation details}
\subsection{The structure of Einstein equations} \label{eq_structure}
The structure of the Einstein field equations in hyperboloidal formalism can be clarified with the help of the ADM 3+1 formulation.

Define the future directed unit normal covector of the $T$-constant hypersurface $\Sigma$
\begin{equation}
    n_{\mu}=
    -\frac{1}{\sqrt{-g^{TT}}}(\mathrm{d}T)_{\mu},
\end{equation}
and the induced metric on $\Sigma$
\begin{equation}
    \gamma_{\mu\nu}=g_{\mu\nu} + n_{\mu}n_{\nu}.
\end{equation}

Eq.(\ref{chi_evo}) is equivalent to the $Z$ component of momentum constraint
\begin{equation} \label{momentum-Z}
\begin{aligned}
\mathcal{M}_Z & = 
\mathcal{G}^{\mu\nu}n_{\mu}\gamma_{\nu Z}
    \\
    & = -\frac{L^3}{Z^3}e^{-\frac{3}{2}\chi}(H\mathcal{G}^{TT}+\mathcal{G}^{TZ})
    \\
    & = \frac{1}{L}e^{\chi/2}
    \left(
    \dot{\chi}- 2H\chi ^{\prime}-4Z\mathrm{Re}\left[ \zeta \bar{\psi}^{\prime} \right]
    \right)
    \\
    & = \frac{1}{L}e^{\chi/2}
    \tilde{\mathcal{M}}_Z
    \\
    & = 0,
\end{aligned}
\end{equation}
where $\mathcal{G}^{\mu\nu} = G_{\mu\nu} - T_{\mu\nu},G_{\mu\nu}=R_{\mu \nu}-\frac{1}{2}Rg_{\mu \nu}$ and $T_{\mu \nu}=2(T_{\mu \nu}^{\psi}+T_{\mu \nu}^{A})-\Lambda g_{\mu \nu}$. On the third and fourth lines, we rewrite eq.(\ref{chi_evo}) and call it $\tilde{\mathcal{M}}_Z$ for convenience.

One can obtain that eq.(\ref{constraint_E}), called $\tilde{\mathcal{H}}$, is exactly
\begin{equation}\label{Hamilton}
    \begin{aligned}
    \tilde{\mathcal{H}}
    & =\frac{L^2}{Z^2}e^{-\chi}\mathcal{H}
     + \frac{L}{Z}He^{-\chi/2}\mathcal{M}_Z
     \\
     & = \frac{L^4}{Z^4}e^{-2\chi}
     \left[
     (1-H^2)\mathcal{G}^{TT}
     - H\mathcal{G}^{TZ}
     \right]
     \\ & =0,
    \end{aligned}
\end{equation}
where the Hamiltonian constraint is defined as
\begin{equation}
    \mathcal{H}  = \mathcal{G}^{\mu\nu} n_{\mu}n_{\nu} 
    = \frac{L^2}{Z^2}e^{-\chi}\mathcal{G}^{TT}.
\end{equation}
The evolutionary equation (\ref{H_evo}) of $H$, called $\tilde{H}$, is exactly
\begin{equation}\label{dotH}
\begin{aligned}
    \tilde{H} &=
    \frac{Z}{2}
    \mathcal{G}^{\mu\nu}
    \gamma_{\mu Z}\gamma_{\nu Z}
    -
    \frac{L^2}{2Z}e^{-\chi}\mathcal{H}
    \\
    & = - \frac{L^4}{2Z^3}e^{-2\chi}
    [\mathcal{G}^{TT}-(1+H^2)(H\mathcal{G}^{TZ}+\mathcal{G}^{ZZ})]
    \\
    & = 0. 
\end{aligned}
\end{equation}

The last Einstein equation including temporal derivative of $\dot{\chi}$ is 
\begin{widetext}
\begin{equation} \label{dotchi}
\begin{aligned}
    \mathcal{G}^{\theta\theta}
      & = 
      \mathcal{G}^{\mu\nu}
      \gamma_{\mu}{}^\theta
      \gamma_{\nu}{}^\theta
      \\
      & = \frac{L^2e^{\chi}}{2} 
      \left\{
\ddot{\chi}-\frac{4}{Z}\dot{H}-\chi ^{\prime}\dot{H}- H^{\prime}\dot{\chi}
+2\dot{H}^{\prime}-2H\dot{\chi}^{\prime}
-\left[ \left( 1-H^2 \right) \chi ^{\prime} \right] ^{\prime}+Z\left[ Z^2\left( \frac{1-H^2}{Z^3} \right) ^{\prime} \right] ^{\prime}
\right. \\
&\left.+\frac{2L^2}{Z^2}e^{-\chi}\left[ \Lambda +2\mu ^2|\psi |^2 \right] -\frac{2Z^2e^{-\chi}B^2}{L^2}+4\left( |\psi ^{\prime}|^2-|\zeta |^2 \right) 
      \right\}
      \\
      & =0.
\end{aligned}
\end{equation}

Using the Bianchi identity of the Einstein tensor $G^{\mu\nu}$ and the covariant conservation of the energy-momentum tensor $T^{\mu\nu}$, we can get
\begin{equation} \label{Bianchi}
    0=\nabla_\mu\mathcal{G}^{\mu\nu}.
\end{equation}

Note that we can write $\{\mathcal{G}^{TT},\mathcal{G}^{TZ},\mathcal{G}^{ZZ}\}$ as linear combinations of $\{\tilde{\mathcal{M}_Z},\tilde{\mathcal{H}},\tilde{H}\}$ using eq.(\ref{momentum-Z}),(\ref{Hamilton}) and (\ref{dotH})
\begin{equation} \label{Bian-elimination}
    \begin{pmatrix}
    &\mathcal{G}^{TT}\\
    &\mathcal{G}^{TZ}\\
    &\mathcal{G}^{ZZ}\\
    \end{pmatrix}
    =
    \frac{Z^3}{L^4}e^{2\chi}
    \begin{pmatrix}
 -H & Z & 0 \\
 H^2-1 & -H Z & 0 \\
 -\frac{H^5}{H^2+1} & \frac{\left(H^4+H^2+1\right) Z}{H^2+1} & \frac{2}{H^2+1} \\
\end{pmatrix}
    \begin{pmatrix}
    &\tilde{\mathcal{M}_Z}\\
    &\tilde{\mathcal{H}}\\
    &\tilde{H}\\
    \end{pmatrix}.
\end{equation}

By substituting eq.(\ref{Bian-elimination}) into $0=-(1-H^2)\nabla_\mu\mathcal{G}^{\mu T}+H\nabla_\mu\mathcal{G}^{\mu Z}$ and setting $\tilde{\mathcal{M}_Z}=\tilde{H}=0$ one can obtain\footnote{Here, we have used $\nabla_\mu\mathcal{G}^{\mu \theta}=0$, which gives $\mathcal{G}^{\theta \theta}=\sin^2\theta\mathcal{G}^{\phi \phi}.$}
  \begin{equation}
    \dot{\tilde{\mathcal{H}}} =
    \frac{H \left(H^2+2\right)}{H^2+1}
    \tilde{\mathcal{H}}^\prime
    +
\left[ \frac{\left( H^4+H^2+2 \right) H^{\prime}}{\left( H^2+1 \right) ^2}-\frac{H\left( HZ\dot{\chi}+4H^2+8 \right)}{2\left( H^2+1 \right) Z} \right] 
\tilde{\mathcal{H}}.
\end{equation}  
\end{widetext}
Thus, the constraint $\tilde{\mathcal{H}}=0$, i.e. eq.(\ref{constraint_E}), will be preserved all the time, provided that it is satisfied on the initial time slice.

Using $0=H\nabla_\mu\mathcal{G}^{\mu T}+\nabla_\mu\mathcal{G}^{\mu Z}$ and setting $\tilde{\mathcal{M}_Z}=\tilde{\mathcal{H}}=\tilde{H}=0$, one can obtain 
\begin{equation} \label{H}
   \frac{2 L^2 e^{\chi } \mathcal{G}^{\theta \theta }}{Z} = 0.
\end{equation}
That is, eq.(\ref{dotchi}) is redundant, which can be derived from eq.(\ref{chi_evo}), (\ref{H_evo}) and (\ref{constraint_E}).
\subsection{The construction of the initial data} \label{construction-H_init}
Integrating eq.(\ref{H_init}), we have
\begin{equation} \label{App_H2}
    H^2=1-C(\tilde{Z})Z^3-Z^3\int_{\tilde{Z}}^Z{F\left( z \right) \mathrm{d}z},
\end{equation}
where $F(Z)$ is the right hand side of eq.(\ref{H_init}) and $C(\tilde{Z})$ is an undetermined integration constant dependent on the lower limit $\tilde{Z}$ of integral. 

Similar to Sec.\ref{RNdS_solution}, we need to find the zero $Z_0$ of $H$ to the square root of eq.(\ref{App_H2}). To do this we substitute $Z_0$ into eq.(\ref{App_H2}) and its derivative with respect to $Z$, and get
\begin{equation} \label{App_Z0}
	\frac{1}{Z_{0}^{3}}+\frac{Z_0}{3}F\left( Z_0 \right) =0,
\end{equation}
\begin{equation}
    C(\tilde{Z}) =-\int_{\tilde{Z}}^{Z_0}{F\left( z \right) \mathrm{d}z}-\frac{Z_0}{3}F\left( Z_0 \right).
\end{equation}
One can solve $Z_0$ numerically via eq.(\ref{App_Z0}), so that  
\begin{equation}\label{App_H}
H\left( Z \right) =\left\{ \begin{array}{c}
	+\sqrt{1-C(\tilde{Z})Z^3-Z^3\int_{\tilde{Z}}^Z{F\left( z \right) \mathrm{d}z}}, Z<Z_0\\
	-\sqrt{1-C(\tilde{Z})Z^3-Z^3\int_{\tilde{Z}}^Z{F\left( z \right) \mathrm{d}z}}, Z>Z_0\\
\end{array} \right. ,
\end{equation}
which could ensure $H\left( Z_c \right) =1,H\left( Z_h \right) =-1$. Note that $\tilde{Z}$ can be set freely.

To set the simulation domain $[Z_1,Z_2]$, we need to solve the initial positions of $Z_c$ and $Z_h$, which can be obtained by substituting $\tilde{Z}=Z_c$ and $\tilde{Z}=Z_h$ into eq.(\ref{App_H}) respectively. That is, both the initial $Z_c$ and $Z_h$ are the roots of the equation $C(\tilde{Z})=0$. 
\section{Physical quantities} \label{Physical quantities}
We monitor the evolution of several physical quantities during our simulation: the scalar field energy $E_\psi$, scalar field charge $Q_\psi$, black hole charge $Q_h$ the irreducible mass $M_{\text{irr}}$ and the rescaled Misner-Sharp mass $M_{\text{MS}}$.

The scalar field energy is calculated as follows:
\begin{equation}
\begin{aligned}
    E_\psi &=-\frac{1}{4\pi}\int_{Z_h}^{Z_c}\sqrt{\gamma}\mathrm{d}^3x  T^{\psi}_{\mu\nu}n^{\mu}n^{\nu}
    \\
    &=
\int_{Z_c}^{Z_h}{\mathrm{d}Z}\frac{2L^5}{Z^3}e^{-\chi /2}\left[ \mu ^2|\psi |^2+\frac{Z^2}{L^2}e^{\chi}\left( |\zeta |^2+|\psi ^{\prime}|^2 \right) \right].
\end{aligned}
\end{equation}
Note that the scalar field energy density is nonnegative, satisfying the weak energy condition.

The scalar field charge is associated with the Nother charge for the conserved current $j^\nu$:
\begin{equation}
\begin{aligned}
    Q_\psi &= 
-\frac{1}{4\pi}\int_{Z_h}^{Z_c}{\mathrm{d}^3x\sqrt{\gamma}}n_{\nu}j^{\nu}
\\
&=-\int_{Z_h}^{Z_c}{\mathrm{d}Z}\frac{2qL^4}{Z^2}\mathrm{Im}\left[ \bar{\zeta} {\psi} \right] .
\end{aligned}
\end{equation}

The black hole charge is evaluated by
\begin{equation} \label{BH-charge}
    Q_h = -\frac{1}{4\pi}\oint_{Z_h}*F
        = - (BL^2) |_{Z_h}
,
\end{equation}
where $*$ represents the Hodge star operation.
The total charge of the spacetime, in the hyperboloidal coordinates, can be evaluated at the future cosmological horizon straightforwardly, $Q_{\text{tot}} = - (BL^2) |_{Z_c}.$

The irreducible mass is $M_{\text{irr}}=\sqrt{S_h/16\pi}$, where $S_h=4\pi L^4/Z_h^2$ represents the apparent horizon area.

The mass of the BH can be calculated as
\begin{equation}
    M_{\text{B}} = M_{\text{irr}} + \frac{Q_h^2}{4 M_{\text{irr}}} - \frac{4}{3}\Lambda M_{\text{irr}}^3.
\end{equation}
This formula is valid for the initial and final state of evolutions where the spacetime region near the apparent horizon is almost static.

The rescaled Misner-Sharp mass is defined as \cite{Maeda2012}
\begin{equation} \label{mass_MS}
    M_{\text{MS}}=\frac{r}{2}(1-g^{\mu\nu}\partial_{\mu}r\partial_{\nu}r-\frac{{\Lambda}r^2}{3}),\ r=\frac{L^2}{Z},
\end{equation}
and the total mass of the spacetime is evaluated at the future cosmological horizon $Z_c$.

\section{Numerical error and convergence}
\label{err}
As mentioned in \ref{eqs}, eq.(\ref{constraint_E}) and (\ref{Maxwell_Z}) are used as the constraint equations to check the validity of our numerics. We have checked that all of them remain satisfied throughout our evolution. 

\begin{figure*}
    \includegraphics[width=8.1cm]{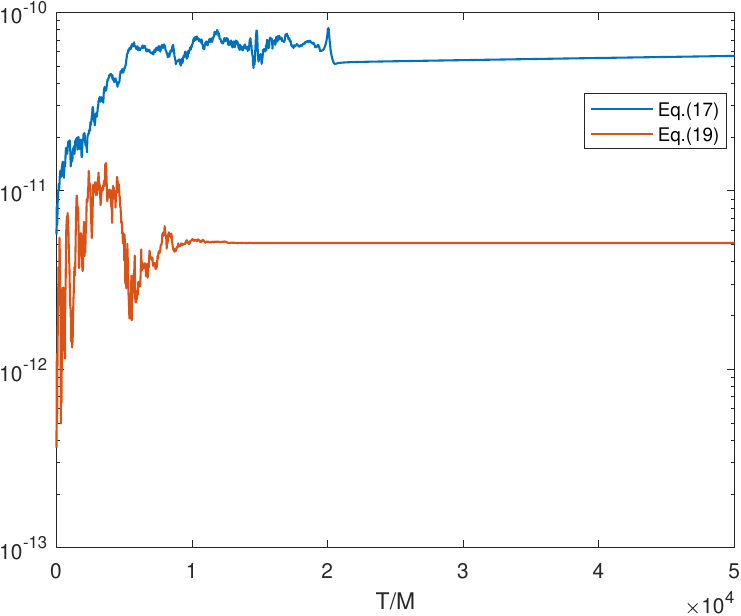}
    \includegraphics[width=8.8cm]{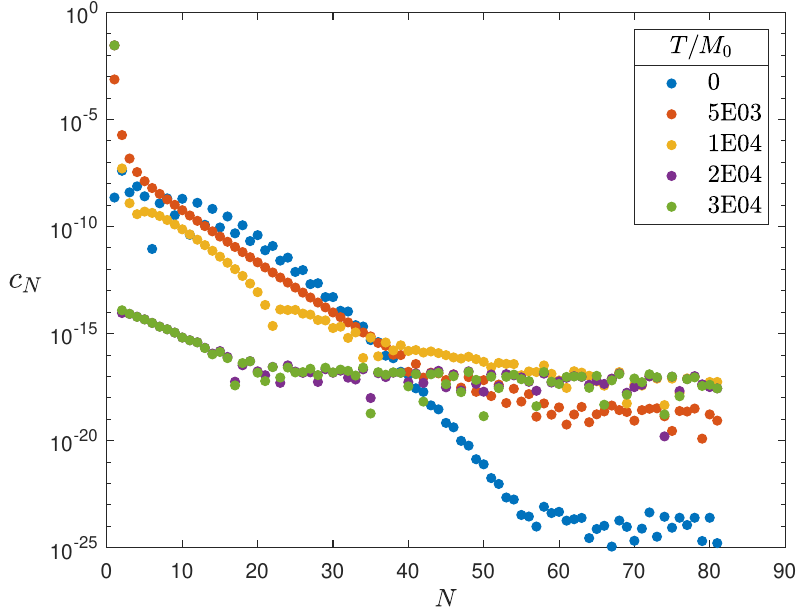}
    \caption{
    Left panel: The variation of the maximal constraint violations along the $Z$ direction with time.
    Right panel: The snapshots of the spectral coefficients $c_N$ of $\chi$. We take $N=81$ points in the $Z$ direction.
    The parameters are $M_0=0.2, Q_0=0.3M_0, \Lambda M_0^2=10^{-3}, qQ_0=0.55, \mu M_0=2\times10^{-4}, k=10^{-4}$. }
    \label{constr&converg}
\end{figure*}

To be specific, we calculate the relative variation $\tilde{\mathcal{H}}/||\tilde{\mathcal{H}}||$ for eq.(\ref{constraint_E}), where $\tilde{\mathcal{H}}$ is defined in \ref{Hamilton} and $||\tilde{\mathcal{H}}||$ represents the sum of the absolute value of each term in eq.(\ref{constraint_E}), i.e. 
\begin{widetext}
\begin{equation}
||\tilde{\mathcal{H}}|| :=
(1+H^2)|\chi ^{\prime}|+2Z[|\zeta |^2+|\psi ^{\prime}|^2+|2H\mathrm{Re}\left( \zeta \bar{\psi}^{\prime} \right) |]
+Z^3\left|(\frac{1-H^2}{Z^3})^{\prime}\right|
+\frac{Ze^{-\chi}}{L^2}+\frac{Z^3e^{-\chi}B^2}{L^2}+\frac{e^{-\chi}L^2}{Z}\left[ \Lambda +2\mu ^2|\psi |^2 \right] .
\end{equation}
\end{widetext}
But the calculation of the relative variation is not applicable to eq.(\ref{Maxwell_Z}), because each term in eq.(\ref{Maxwell_Z}) is very close to zero in the early and late stages of evolution. We straightly calculate the absolute value of right hand side for eq.(\ref{Maxwell_Z}). We show the variation of the maximal constraint violations along the $Z$ direction with time in Fig.\ref{constr&converg}.

We have also checked the spectral convergence of our numerical simulation by examining spectral coefficients $c_N$ for each dynamical variable. We demonstrate such a spectral convergence for $\chi$ in Fig.\ref{constr&converg}, where one can see the exponential convergence.

\bibliography{RNdS}
\end{document}